\documentclass[secnum]{kokyu}%%% eqno resets every section
\usepackage{amssymb, amsmath}%%% option packages
\usepackage{graphics}%%% option packages
\usepackage[dvips]{graphicx}%%% option packages
\usepackage{eepic}%%% option packages
\usepackage{epic}%%% option packages
\newtheorem{thm}{Theorem}[section]

\theoremstyle{definition}

\theoremstyle{remark}
\title{ On Three Imaginary-time Path Integral Formulas\\
 with Magnetic Fields\\ in Relativistic Quantum Mechanics
\thanks{based on talk at RIMS Joint Research  
{\sl ``Introductory Workshop on Path Integrals
  and Pseudo-Differential Operators"}, October 7--10, 2014.}}
\author{Takashi \textsc{Ichinose} \footnote{Professor emeritus,
Kanazawa University, Kanazawa 920-1192, Japan.}}
\AuthorHead{Takashi Ichinose} %optional but recommended
\TitleHead{Three Imaginary-time Path Integral Formulas in Relativistic Quantum
Mechanics}  
    %optional
\begin{document}
%
% The text goes here.
% Be sure to use the appropriate "theorem-like" environment as
% is the following examples.  Never use plain TeX commands for these, as
% they will cause interference with the styles of other papers.

\maketitle

%\tableofcontents      %optional

\begin{abstract}      %optional
Three magnetic relativistic Schr\"odinger
operators are considered, corresponding to the classical relativistic
Hamiltonian symbol with both magnetic vector and electric scalar potentials.
Path integral representations
for the solutions of their respective imaginary-time relativistic Schr\"odinger
equations, i.e. heat equations are given in two ways. The one is by means of
the probability path space measure coming from the L\'evy process concerned,
and the other is through time-sliced approximation with Chernoff's theorem.
\end{abstract}

\noindent
Table of contents

\medskip
{\footnotesize
\noindent
1. Introduction

\noindent
2. Three magnetic relativistic Schr\"odinger operators

\noindent
3. Imaginary-time path integral formulas for magnetic relativistic
Schr\"odinger operators

\noindent
$\,\,\,\,$3.1. The case for the Weyl pseudo-differential operator $H_A^{(1)}+V$

\noindent
$\,\,\,\,$3.2. The case for the modified Weyl pseudo-differential operator $H_A^{(2)}+V$

\noindent
$\,\,\,\,$3.3. The case for $H_A^{(3)}+V = \sqrt{(-i\nabla-A)^2+m^2} +V$ with square
root

\noindent
$\,\,\,\,$3.4. Summary of three path integral formulas

\noindent
$\,\,\,\,$3.5. Path integral formulas (3.5), (3.12) and (3.15) as time-sliced 
                 approximation

\noindent
4. Some observation on
Chernoff's theorem and path integral by time-sliced approximation

\noindent
$\,\,\,\,$4.1. Time-sliced approximation in strong topology

4.1.1. Schr\"odinger operator with scalar potential $V(x)$

4.1.2. Schr\"odinger operator with vector and scalar potentials
$A(x)$ and $V(x)$

4.1.3. Dirac operator with vector and scalar potentials $A(x)$ and $V(x)$

\noindent
$\,\,\,\,$4.2. Time-sliced approximation in norm and pointwise

4.2.1. Trotter--Kato product formula and Chernoff's theorem
in norm

\newpage
4.2.2. Time-sliced approximation for Schr\"odinger equation in real
and imaginary time 
\newline \indent\qquad --- convergence in norm and pointwise

\noindent
References

%{}
}

%%%%%%%%%%%%%%%
\section{Introduction} %sect.1

{\it Path integral} is a marvelous idea invented
by R. P. Feynman ([F-48], [F-05], [FH-65]. cf. [D-33,35], [D-45])
to give a practically very useful and now figurative sublimed way
to write down the solution of the {\it real-time} Schr\"odinger
equation in nonrelativistic quantum mechanics.

In this paper, we deal with the problem in relativistic quantum mechanics
to consider the relativistic Schr\"odinger equation in {\it imaginary time}.
In the literature there are 3 kinds of relativistic Schr\"odinger operators
$H = H_A + V$ for a spinless particle of mass $m\geq 0$ corresponding to
%%%
\begin{equation}
\sqrt{(\xi-A(x))^2 +m^2} +V(x)\,, \qquad
(\xi,x) \in {\bf R}^d \times {\bf R}^d,
\end{equation}%(1.1)
under {magnetic} vector potential $A(x)$ and electric scalar potential
$V(x)$, depending on {\sl how to quantize
the kinetic energy term} $\sqrt{(\xi-A(x))^2 +m^2}$.
This $H$ is used in the situation where we may ignore QFT effect like
particles creation and annihilation,  but should take relativistic
effect into consideration.

We give three path integral representation formulas for the solutions
for their respective imaginary-time relativistic Schr\"odinger equations,
i.e. heat equations, by means of
the probability path space measure coming from L\'evy process concerned.
We also discuss the path integral by time-sliced approximation. It is well-known
that this method also can plainly give a meaning for Schr\"odinger equation
by the Trotter--Kato product formula, if the Schr\"odinger operator
has only electric scalar potential. But if it has also
{\it magnetic vector potential}, we should use Chernoff's theorem instead.
This wisdom also applies to Dirac equation.

This paper is of expository character, having in sections 2 and 3
description and content which overlap with  another
a little more elaborate paper on the subject in
{\sl RIMS Kyoto Univ. K\^oky\^uroku} {\bf 1797}(2012) [I-12a].
Their detailed version was in the meanwhile published in my
paper [I-13] and also brief note [I-12b].
So I would not like to repeat the whole story here but only to write
a short survey describing the points how to obtain the three path integral
representation formulas with sketch of proof. In \S4, the path integral
by time-sliced approximation is further studied for some other evolution
equations in quantum mechanics in real and imaginary time by means of
 {\it Chernoff's theorem}, also discussing its convergence, not only in
strong topology, but also in operator norm and pointwise for the integral kernels.
The content is almost independent of the three relativistic Schr\"odinger operators
up to the previous section. The observation of this last section might contain
something new.

For the reader's convenience, the table of contents of [I-13] is as follows:

%%%%%%%%%%%%%%%%%%%%%%%%%%%%%%%
\medskip\noindent{\footnotesize
%Table of contents

\noindent
1. Introduction;

\noindent
2. Three magnetic relativistic Schr\"odinger operators:
2.1.Their definition and difference;
2.2.Gauge-covariant or not;

\noindent
3. More general definition of magnetic relativistic Schr\"odinger
   operators and their selfadjointness:
3.1.The most general definition of $H_A^{(1)}$, $H_A^{(2)}$ and $H_A^{(3)}$;
3.2.Selfadjointness with negative scalar potentials;

\noindent
4. Imaginary-time path integrals for magnetic relativistic Schr\"odinger operators:
4.1.Feynman--Kac--It\^o type formulas for magnetic relativistic
Schr\"odinger operators;
4.2.Heuristic derivation of path integral formulas;

\noindent
5. Summary

{}
}

%%%%%%%%%%%%%%%%%%%%%%%%%%%%%%%%
\section{Three magnetic relativistic Schr\"odinger operators} %sect.2

The three relativistic Schr\"odinger operators concerned are the following.
The first one is the Weyl pseudo-differential operator defined
through mid-point prescription
 $\, H^{(1)} :=  H^{(1)}_A + V$ considered by Ichinose and Tamura
([IT-86], [I-89, 95], [NaU-90]), the second $H^{(2)} :=  H^{(2)}_A + V$
the modification of the first one by
Iftimie, M\u{a}ntoiu and R.~Purice [IfMP-07, 08, 10], and
the third $H^{(3)} :=  H^{(3)}_A + V$ defined with the square root $H^{(3)}_A$ of
the nonnegative selfadjoint operator $(-i\nabla-A(x))^2+m^2$.

For simplicity, assume that {\it $A(x)$ is smooth and $V(x)$ bounded below}.

\bigskip\noindent
{\bf (1)} {\sl Weyl pseudo-differential operator
$\, H^{(1)} :=  H^{(1)}_A + V$} (e.g. [IT-86; I-89, 95]) with
\begin{eqnarray}%(2.1)
(H^{(1)}_A f)(x)
:&\!=& \tfrac1{(2\pi)^{d}} \int\!\!\int_{{\bf R}^d\times {\bf R}^d}
   e^{i(x-y)\cdot\xi}
 \small{\sqrt{\Big(\xi-A\big(\tfrac{x+y}{2}\big)\Big)^2 +m^2}}\,
   f(y) dyd\xi \\
&\!=& \tfrac1{(2\pi)^{d}} \int\!\!\int_{{\bf R}^d\times {\bf R}^d}\,
   e^{i(x-y)\cdot (\xi+A(\tfrac{x+y}{2}))}
  \sqrt{\xi^2 +m^2} f(y) dyd\xi \,.\nonumber
\end{eqnarray}
Here, with
$f \in C_0^{\infty}({\bf R}^d)$ or $f \in {\cal S}({\bf R}^d)$, the
integrals on the right-hand side are oscillatory integrals.

\bigskip\noindent
{\bf (2)} {\sl Modified Weyl pseudo-differential operator
$\, H^{(2)} :=  H^{(2)}_A + V$} [IfMP-07, 08, 10] with
\begin{eqnarray}%(2.2)
(H^{(2)}_Af)(x) %\nonumber\\
:&\!=& \tfrac1{(2\pi)^{d}} \int\!\!\int_{{\bf R}^d\times {\bf R}^d}
   \!\!e^{i(x-y)\cdot\xi}
  \small{\sqrt{\big(\xi-
   \int_0^1 A((1-\theta)x+\theta y)d\theta \big)^2 +m^2}}\,
   f(y) dyd\xi \nonumber\\
&\!=& \tfrac1{(2\pi)^{d}}\! \int\!\!\int_{{\bf R}^d\times {\bf R}^d}
   \!\!e^{i(x-y)\cdot (\xi +\int_0^1 A((1-\theta)x+\theta y)d\theta)}
   \sqrt{\xi^2 +m^2}f(y) dyd\xi
\end{eqnarray}

\bigskip\noindent
{\bf (3)} {\sl $H^{(3)} := H^{(3)}_A + V$ defined with square root
\begin{equation}
H^{(3)}_A := \sqrt{(-i\nabla-A(x))^2+m^2}
\end{equation}
of the nonnegative selfadjoint operator $(-i\nabla-A(x))^2+m^2$}.
This $H^{(3)}$ is used, e.g., to study ``stability of matter"
in relativistic quantum mechanics in Lieb--Seiringer [LSei-10].

%\bibitem[ITa-86]

\bigskip\noindent
{\sl Known facts for $H^{(1)}$, $H^{(2)}$ and $H^{(3)}$}

%\noindent
1$^o.$ With suitable reasonable conditions on $A(x)$ and $V(x) \geq 0$, they
all define {\it selfadjoint} operators  in $L^2({\bf R}^d)$,
which are bounded below.
For instance, they become selfadjoint operators defined as quadratic forms,
for $H^{(1)}$ and $H^{(2)}$, when
$A \in L_{\operatorname loc}^{1+\delta}({\bf R}^d; {\bf C}^d)$
for some $\delta>0$ and $V \in L^1_{\operatorname loc}({\bf R}^d)$
(cf. [I-89, 13], [IfMP-07, 08, 10]), while for $H^{(3)}$,
when $A \in L_{\operatorname loc}^2({\bf R}^d; {\bf C}^d)$ and
$V \in L^1_{\operatorname loc}({\bf R}^d)$
(e.g. [CFKS-87, pp.8--10] or [I-13]).

In fact further, they are bounded below by the {\sl same lower bound},
in particular,
\begin{equation}
 H^{(j)}_A \geq m\,, \qquad j= 1, \,2, \,3.
\end{equation}%(2.4)

%\noindent
2$^o.$ $H_A^{(2)}$ and $H_A^{(3)}$ are covariant under gauge transformation,
i.e. it holds for every
$\varphi \in {\cal S}({\bf R}^d)$ that
$\,\,H^{(j)}_{A+\nabla \varphi} = e^{i\varphi}H^{(j)}_{A}e^{-i\varphi}$,
$j=2,3$. However, $H_A^{(1)}$ is not.

%\noindent
3$^o.$ All these three operators are different in general,
but coincide, if $A(x)$ is
linear in $x$, i.e. if $A(x) = \dot{A}\cdot x\,$ with $\dot{A}: d\times d$
real symmetric {\sl constant matrix}, then $H_A^{(1)} = H_A^{(2)} = H_A^{(3)}$.
So, this holds for uniform magnetic fields with $d=3$.

%%%%%%%%%%%%%%%%%%%%%%%%%%%%%%%%%%%%
\section{Imaginary-time Path integral for
magnetic relativistic Schr\"odinger operators}%\sect3

For each $H= H_A +V$ of the three magnetic relativistic Schr\"odinger operators
$H^{(1)}= H_A^{(1)}+V$, $H^{(2)}= H_A^{(2)}+V$ and $H^{(3)}= H_A^{(3)}+V$,
consider imaginary-time relativistic Schr\"odinger equation
\begin{equation}%(3.1)
\left\{
 \begin{array}{rl}
 \displaystyle{\tfrac{\partial}{\partial t}} u(t,x) &= -[H-m]u(t,x),
                                 \quad t>0,\\
     u(0,x) &= g(x), \quad\qquad\qquad\quad x \in {\bf R}^d.
 \end{array}\right.
\end{equation}

The solution of this Cauchy problem is given by the semigroup
$u(t,x) = (e^{-t[H-m]}g)(x)$. We want to find a path integral formula
for each $e^{-(H^{(j)}-m)}g,\, j=1,2,3$.

\subsection{The case for the Weyl pseudo-differential operator
$H^{(1)}= H^{(1)}_A +V$}
%\subsect3.1

$H^{(1)}_A$, in (2.1),  can be rewritten as an integral operator:
\begin{eqnarray}
([H^{(1)}_A-m]f)(x)
&\!=\!& -\int_{|y|>0} [e^{-iy\cdot A(x+\tfrac{y}{2})} f(x+y) -f(x)
          \nonumber\\
          &&\qquad\qquad-I_{\{|y|<1\}} y\!\cdot\! (\nabla-iA(x))\,f(x)]n(dy)\nonumber\\
&\!=\!& -\lim_{r\downarrow 0} \int_{|y|\geq r}
              [e^{-iy\cdot A(x+\tfrac{y}{2})} f(x+y) -f(x)]\,n(dy)\nonumber\\
&\!=\!& -\, \hbox{\rm p.v.}\!\int_{|y|>0}
              [e^{-iy\cdot A(x+\tfrac{y}{2})} f(x+y) -f(x)]\,n(dy) %#(15),nein(7)
\end{eqnarray}
where {$n(dy) =n(y)dy$} is an $m$-dependent measure on ${\bf R}^d\setminus\{0\}$,
called {\it L\'evy measure}, with density
$$%\begin{equation}
n(y) =\left\{
        \begin{array}{rl}
        2(2\pi)^{-(d+1)/2}m^{d+1}(m|y|)^{-(d+1)/2}
                K_{(d+1)/2}(m|y|), & \quad m>0,\\
        \pi^{-(d+1)/2}\Gamma\big(\tfrac{d+1}{2}\big) |y|^{-(d+1)}, &\quad m=0
       \end{array}\right.
$$%\end{equation}
It appears in the {\it L\'evy--Khinchin formula} :
\begin{equation}%#(8)
 \sqrt{\xi^2+m^2}-m = -\!\int_{|y|>0}
  (e^{i y\cdot \xi}-1-i\xi\! \cdot\! y I_{\{|y|<1\}})n(dy)
= -\! \lim_{r\rightarrow 0+} \int_{|z|\geq r} (e^{i z\cdot \xi}-1)n(dz)
\end{equation}
%\bigskip\noindent
{\sl Proof of (3.2)}.  By the L\'evy--Khinchin formula (3.3),
\begin{eqnarray*}
(H_A^{(1)}f)(x)
&\!=& (2\pi)^{-d}\!\int\!\int\,
    e^{i(x-y)\cdot (\xi+A(\tfrac{x+y}{2}))}
     \Big[m\!-\!\!\! \lim_{r\rightarrow 0+} \int_{|z|\geq r}
     (e^{i z\cdot \xi}-1)n(dz)\Big]
     f(y) dyd\xi\\
&\!=& (2\pi)^{-d}\Big[m\int\int\,e^{i(x-y)\cdot\xi}
   e^{i(x-y)\cdot A(\tfrac{x+y}{2})}\, dyd\xi\\
  &&-\!\! \lim_{r\rightarrow 0+}\int\int\int_{|z|\geq r}\,
    \big(e^{i(x-y+z)\cdot\xi}- e^{i(x-y)\cdot\xi}\big)n(dz)
       e^{i(x-y)\cdot A(\tfrac{x+y}{2})}\, f(y)\, dyd\xi\Big]\\
&\!=&m\!
  \int \delta(x-y)e^{i(x-y)\cdot A(\tfrac{x+y}{2})}f(y)\,dy\\
  &&-\!\lim_{r\rightarrow 0+}\int\int_{|z|\geq r}\,
    \big(\delta(x-y+z)- \delta(x-y)\big)n(dz)
       e^{i(x-y)\cdot A(\tfrac{x+y}{2})}\, f(y)\, dy\\
&\!=& mf(x)
  -\lim_{r\rightarrow 0+}\int\int_{|z|\geq r}\,
    \big(e^{-iz\cdot A(x+\tfrac{z}{2})}f(x+z) - f(x)\big)\, n(dz)
\end{eqnarray*}
\qed

%%%%%%%%%%%%%%%%
\bigskip\noindent
{\sl Some Notations from L\'evy process to represent
$e^{-t[H^{(1)}-m]}g$ by path integral}

For more details, we refer to [IkW-81, 89].

\medskip\noindent
{$\cdot D_x([0,\infty)\rightarrow {\bf R}^d)$} : space of right-continuous paths
$X: \![0,\infty)\!\rightarrow\! {\bf R}^d$ with left-hand limits
(called {\it ``c\`adlag} paths") with $X(0)=\!x$

\medskip\noindent
{$\cdot \lambda_x$} : probability measure on
$D_x([0,\infty)\rightarrow {\bf R}^d)$ such that
\begin{equation}% #(9)
e^{-t[\sqrt{\xi^2+m^2}-m]}
 = \int_{D_x([0,\infty)\rightarrow {\bf R}^d)}
  e^{i (X(t)-x)\cdot\xi}d\lambda_x(X), \quad t\geq 0, \,\, \xi \in {\bf R}^d
\end{equation}%

\medskip\noindent
$\cdot N_X(dsdy)$: {\it counting measure} on
$[0,\infty)\times ({\bf R}^d\setminus \{0\})$
 to count the number of discontinuities of the path $X(\cdot)$, i.e.
$N_X((t,t']\times U) := \#\{s \in (t,t'];\, 0\neq X(s)-X(s-) \in U\}$
 ($0 <t <t', U \subset {\bf R}^d\setminus \{0\}: \hbox{\rm Borel\, set}$).
It satisfies
$\displaystyle{\int_{D_x}}N_X(dsdy)\, d\lambda_x(X) =dsn(dy)$.

\medskip\noindent
$\cdot \widetilde{N}_X(dsdy)$ $ := N_X(dsdy) -dsn(dy)
$

\medskip
Then any path $X \in D_x([0,\infty)\rightarrow {\bf R}^d)$ can be
expressed with
$N_x(\cdot),\, \widetilde{N}_X(\cdot)$ as
$$
X(t) = x+ \int_0^{t+}\int_{|y|\geq 1} y { N_X(dsdy)}
               + \int_0^{t+}\int_{0<|y|<1} y { \widetilde{N}_X(dsdy)}.
$$

\begin{thm} \mbox{\rm [ITa-86, I-95]} %Thm3.1
\begin{eqnarray}
\quad({e^{-t[H^{(1)}-m]}g})(x) &\!=\!&
 \int_{D_x([0,\infty)\rightarrow {\bf R}^d)} e^{-S^{(1)}(t,X)}
  g(X(t))\,d\lambda_x (X),\\ %#(10)
{S^{(1)}(t,X)}&\!=\!& i\int_0^{t+}\int_{|y|\geq 1}
                 A(X(s-)+\tfrac{y}{2})\!\cdot\! y\, { N_X(dsdy)} \nonumber\\
 &&  +i\int_0^{t+}\int_{0<|y|<1}
        A(X(s-)+\tfrac{y}{2})\!\cdot\! y\,{ \widetilde{N}_X(dsdy)}\nonumber\\
 && +i\int_0^{t} ds\, \hbox{\rm p.v.}\! \int_{0<|y|<1}
              A(X(s)+\tfrac{y}{2})\!\cdot\! y\, { n(dy)} +\int_0^t V(X(s))ds.
\nonumber
\end{eqnarray}
\end{thm}
For some recent related result on the mass-zero limit problem with $H^{(1)}$,
see [IM-14].

\bigskip
{\sl Proof (Sketch)}. Let $k_0(t,x-y)$ be the integral kernel of
$e^{-t(\sqrt{-\Delta+m^2}-m)}$, put
\begin{align}%(3.6),(3.7)
(F(t)g)(x) :&\!=\! \displaystyle{\int_{{\bf R}^d}} k_0(t,x-y)
  e^{-iA\big(\tfrac{x+y}{2}\big)\cdot(y-x)-V\big(\tfrac{x+y}{2}\big)t}g(y)dy,\\
\intertext{which can be rewritten as}
(F(t)g)(x) &\!=\! \int_{D_x}
  e^{-iA\big(\tfrac{x+X(t)}{2}\big)\cdot(X(t)-x)
             -V\big(\tfrac{x+X(t)}{2}\big)t}g(X(t))d\lambda_x(X).
\end{align}%
For the definition (3.6), note the second expression of the definition
(2.1) of $H_A^{(1)}$.

Then we do partition of the time interval $[0,t]$ into $n$ small
subintervals with the same width $t/n$:
$0=t_0<t_1< \cdots < t_n=t,\,\, t_j-t_{j-1}=t/n$, and put
\begin{eqnarray}%(3.8)
&&S_n(x_0,\cdots,x_n)
:= i\sum_{j=1}^n A\big(\tfrac{x_{j-1}+x_j}{2}\big)\cdot(x_{j}-x_{j-1})
   +\sum_{j=1}^n V\big(\tfrac{x_{j-1}+x_j}{2}\big)\tfrac{t}{n},{}\quad{}\\
&&x_j:=X(t_j) (j=0,1,2,\dots,n); \,x=x_0:=X(t_0), \,x_n:=X(t_n)\equiv X(t),
\nonumber
\end{eqnarray}%{equation}
where note that the assignment $t_j \mapsto X(t_j)$ is in the reversed time
order.

%%%%%%%%

Substitute these $n+1$ points of path $X(\cdot)$ into $S_n(x_0,\cdots,x_n)$
to get
\begin{eqnarray}%(3.10)
\,\,\,S_n(X):&=& S_n(X(t_0),\cdots,X(t_n))\\
&=& i\sum_{j=1}^n
 A\Big(\tfrac{X(t_{j-1})\!+\!X(t_j)}{2}\Big)\cdot(X(t_{j})\!-\!X(t_{j-1}))
+\sum_{j=1}^n V\Big(\tfrac{X(t_{j-1})\!+\!X(t_j)}{2}\Big)\tfrac{t}{n}.
\nonumber
\end{eqnarray}
Then the $n$ times product of $F(t/n)$ turns out
\begin{eqnarray}%(14)
\big(F({t}/{n})^n g\big)(x)%\nonumber\\
  &=& \overbrace{\int_{{\bf R}^d}\cdots\int_{{\bf R}^d}}
       ^{\mbox{$n$ times}}
   \prod_{j=1}^n k_0(t/n,x_{j-1}-x_j)
    e^{-S_n(x_0,\cdots,x_n)}g(x_n)dx_1\cdots dx_n \nonumber\\
&=&\! \int_{D_x} e^{-S_n(X)}g(X(t))\, d\lambda_x(X)\\%#(14)
&=&\! \int_{D_x} e^{-\footnotesize{i\sum_{j=1}^n
 A\Big(\tfrac{X(t_{j-1})\!+\!X(t_j)}{2}\Big)\cdot(X(t_{j})\!-\!X(t_{j-1}))
-\sum_{j=1}^n V\Big(\tfrac{X(t_{j-1})\!+\!X(t_j)}{2}\Big)\tfrac{t}{n}}}
\nonumber\\
&&\qquad\qquad\qquad\qquad\qquad\qquad\qquad\qquad\qquad\quad
                             \times g(X(t))\, d\lambda_x(X).\nonumber
\end{eqnarray}
We have to show convergence of each side of (3.10).

We shall use {\it Chernoff's theorem} for the left-hand side (LHS),
while {\it It\^o formula} for the right-hand side (RHS).

\bigskip
{\sl Proof of convergence of $LHS$ of (3.10)}. We need

\bigskip\noindent
{\bf Lemma A}.
 $\quad \displaystyle{F({t}/{n})^n g \rightarrow e^{-t[H^{(1)}-m]}g\,\,\, \,
\hbox{\rm in}\,\, \,L^2({\bf R}^d), \quad n\rightarrow \infty}$.

\bigskip
The proof of Lemma A is essentially an application of the Chernoff's theorem,
although it was proved directly in [ITa-86] or [I-13].
Note here that if the vector potential $A(x)$ is present, one cannot use the
Trotter--Kato product formula instead of the Chernoff's theorem.

\medskip\noindent
{\bf Chernoff's Theorem}. [Ch-74]
$\,\,\,${\it Let $F$ be a strongly continuous function on $[0,\infty)$
 with values in the Banach space ${\cal L}({\bf X})$
of bounded linear operators on a Banach space ${\bf X}$.
Assume that $F$ further satisfies the following conditions}:
(i) {\it $F(0) = I$ ($I$: identity operator on ${\bf X}$), and there exists
a real $a$ such that $\|F(t)\| \leq e^{at}$  for all $t\geq 0$;}

\noindent
(ii) {\it The linear operator $F'(0)\upharpoonright_{D[F'(0)]}$ is closable,
and the closure $\overline{F'(0)} := L$
      generates a strongly continuous semigroup $e^{-tL} $.}

{\it Then
$F(t/n)^n$ converges to $e^{-tL}$  strongly, as $n\rightarrow\infty$,
   uniformly on each finite interval in  $t \geq 0$. }

\bigskip
Note that condition (ii) means nothing but that $u(t) :=e^{-tL}u_0$ is the
solution of equation $\tfrac{d}{dt}u(t) = -Lu(t)$ with initial data $u(0)=u_0$.
In \S 4, we shall give some observation on Chernoff's theorem as to how useful
it makes sense to path integral by time-sliced approximation.

\medskip
Now, for the proof of Lemma A, we content ourselves with only confirming
applicability of the Chernoff's theorem on ${\bf X} = L^2({\bf R}^d)$ with
$L= H^{(1)}\!\!-\!\!m$, and (3.6), i.e.
\begin{equation}
(F(t)g)(x) := \int_{{\bf R}^d} (e^{-t[\sqrt{-\Delta+m^2}-m]})(x-y)
      e^{iA(\tfrac{x+y}2)(y-x)-V(\tfrac{x+y}2)t}g(y)dy,
\end{equation}
where we are writing the integral kernel $k_0(t,x-y)$ of the semigroup
$e^{-t[\sqrt{-\Delta+m^2}-m]}$ as $(e^{-t[\sqrt{-\Delta+m^2}-m]})(x-y)$.

Indeed, we can show that
  $\tfrac{I-F(t)}{t} \rightarrow H^{(1)}$ in strong resolvent sense
as $t \downarrow 0$,
which yields Lemma A, namely, that LHS of (3.10) converges to
$e^{-t[H^{(1)}-m]}g$ as $n\rightarrow \infty $. \qed

\bigskip
{\sl Proof of convergence of RHS of (3.10)}. We are going to show
\begin{eqnarray*}
\hbox{\rm RHS}\,\, \hbox{\rm of}\,\, \hbox{\rm  (3.10)}
&=& \int_{D_x} e^{-S_n(X)}g(X(t))\, d\lambda_x(X)\\
&=&\int_{D_x} e^{-\footnotesize{i\sum_{j=1}^n
 A\Big(\tfrac{X(t_{j-1})\!+\!X(t_j)}{2}\Big)\cdot(X(t_{j})\!-\!X(t_{j-1}))
-\sum_{j=1}^n V\Big(\tfrac{X(t_{j-1})\!+\!X(t_j)}{2}\Big)\tfrac{t}{n}}}\\
&&\qquad\qquad\qquad\qquad\qquad\qquad\qquad\qquad\qquad
             \times g(X(t))\, d\lambda_x(X)\\
&\rightarrow& \int_{D_x} e^{-S(X)}g(X(t))\, d\lambda_x(X),
\qquad\hbox{\rm as}\,\, n\rightarrow \infty.
\end{eqnarray*}

%%%%%%%%%%
In fact, in equation (3.8),
we can use {\it It\^o's formula} [IkW-81,89] for the $j$-th summand of
the first term on the right
to rewrite it as a sum of three integrals on the $t$-interval
$\,t_{j-1}\leq s <t_j$:
{\footnotesize
\begin{eqnarray*}
&&A\big(\tfrac{X(t_{j-1})+X(t_j)}{2}\big)\cdot\big(X(t_{j})-X(t_{j-1})\big)\\
&=& \int_{t_{j-1}}^{t_j+} \int_{|y|>0}
  \Big[A\big(\tfrac{X(s-)+X(t_{j-1})+yI_{|y|\geq 1}(y)}{2}\big)
      \cdot\big(X(s-)-X(t_{j-1})+yI_{|y|\geq  1}(y)\big)\\
 &&\qquad\quad - A\big(\tfrac{X(s-)+X(t_{j-1})}{2}\big)
  \cdot\big(X(s-)-X(t_{j-1})\big)\Big]\,N_X(dsdy)\\
&&+ \int_{t_{j-1}}^{t_j+}\int_{|y|>0}
\Big[A\big(\tfrac{X(s-)+X(t_{j-1})+yI_{|y|< 1}(y)}{2}\big)
      \cdot\big(X(s-)-X(t_{j-1})+yI_{|y|< 1}(y)\big)\\
 &&\qquad\quad
 - A\big(\tfrac{X(s-)+X(t_{j-1})}{2}\big)\cdot\big(X(s-)-X(t_{j-1})\big)
    \Big] \,\widetilde{N}(dsdy)\\
&&+\int_{t_{j-1}}^{t_j}\int_{|y|>0}
  \Big[A\big(\tfrac{X(s)+X(t_{j-1})+yI_{|y|< 1}(y)}{2}\big)
    \cdot\big(X(s)-X(t_{j-1})+yI_{|y|< 1}(y)\big)\\
 &&\qquad\quad - A\big(\tfrac{X(s)+X(t_{j-1})}{2}\big)
    \cdot\big(X(s)-X(t_{j-1})\big)\\
 &&\,-I_{|y|<1}(y)\Big(\big(\tfrac12(y\cdot\nabla)A\big)
 \big(\tfrac{X(s)+X(t_{j-1}}{2}\big) \cdot\big(X(s)-X(t_{j-1})\big)
 + y\cdot A\big(\tfrac{X(s)+X(t_{j-1})}{2}\big)\Big)\Big] \,dsn(dy)
\end{eqnarray*}
}
It follows that
{\small
\begin{eqnarray*}
S_n(X) &=&  i\sum_{j=1}^n
  A\big(\tfrac{X(t_{j-1})\!+\!X(t_j)}{2}\big)\cdot(X(t_{j})\!-\!X(t_{j-1}))
+\sum_{j=1}^n V\big(\tfrac{X(t_{j-1})\!+\!X(t_j)}{2}\big)\tfrac{t}{n}\\
&=& \sum_{j=1}^n\Big[i\int_{t_{j-1}}^{t_j+} \int_{|y|>0}\cdots  N_X(dsdy)
 +i\int_{t_{j-1}}^{t_j+}\int_{|y|>0}\cdots \widetilde{N}(dsdy)\\
 &&\qquad\qquad\qquad\qquad
   + i\int_{t_{j-1}}^{t_j}\int_{|y|>0}\cdots dsn(dy)
  + V(\tfrac{X(t_{j-1}+X(t_j)}2)\tfrac{t}{n}\Big],
\end{eqnarray*}
}
which, as $n\rightarrow \infty$, converges to
{\small
\begin{eqnarray*}
&&i\Big[\int_0^{t+}\int_{|y|\geq 1}
                 A(X(s-)+\tfrac{y}{2})\!\cdot\! y\, N_X(dsdy)
 \quad +\int_0^{t+}\int_{0<|y|<1}
            A(X(s-)+\tfrac{y}{2})\!\cdot\! y\,\widetilde{N}_X(dsdy)\\
 &&\qquad\qquad\qquad
  +\int_0^{t} ds\, \hbox{\rm p.v.}\! \int_{0<|y|<1}
                 A(X(s)+\tfrac{y}{2})\!\cdot\! y\, n(dy)\Big]
  +\int_0^t V(X(s))ds\\
&\equiv& S^{(1)}(t,X),
\end{eqnarray*}
}
whence
$$
\hbox{\rm RHS\, of\, (3.10) }
= \int_{D_x} e^{-S_n(X)}g(X(t))\, d\lambda_x(X)
\rightarrow \int_{D_x} e^{-S^{(1)}(t,X)}g(X(t))\, d\lambda_x(X).
$$
\qed

%%%%%%%%%%%%%%%%%%%%%%%%

\subsection{The case for the Weyl pseudo-differential operator
  modified by Iftimie--M\u{a}ntoiu--Purice
  $\, H^{(2)} :=  H^{(2)}_A + V$}%subsect3.2

First note that we can rewrite $H^{(2)}_A$, in (2.2),
similarly for $H^{(1)}_A$, as integral operator
\begin{eqnarray}
([H^{(2)}_A-m]f)(x)
&=& -\int_{|y|>0} [e^{-iy\cdot \int_0^1 A(x+\theta y)d\theta}f(x+y) -f(x)
  \\
 &&\qquad\qquad-I_{\{|y|<1\}} y\!\cdot\! (\nabla-iA(x))\,f(x)]n(dy)\nonumber\\
&\!=& -\lim_{r\downarrow 0} \int_{|y|\geq r}
              [e^{-iy\cdot \int_0^1 A(x+\theta y)d\theta}f(x+y) -f(x)]
                  \,n(dy)\nonumber\\
&\!=& -\, \hbox{\rm p.v.}\!\int_{|y|>0}
              [e^{-iy\cdot \int_0^1 A(x+\theta y)d\theta}f(x+y) -f(x)]\,n(dy).
\nonumber
\end{eqnarray}%

\bigskip\noindent
\begin{thm} \mbox{\rm [IfMP-07, 08, 10]}%Thm3.2
\begin{eqnarray}
(e^{-t[H^{(2)}-m]}g)(x) &\!=\!&
  \int_{D_x([0,\infty)\rightarrow {\bf R}^d)} e^{-S^{(2)}}(t,X)
        g(X(t))\,d\lambda_x (X),\\ %#(17)
S^{(2)}(t,X) &\!=\!& i\int_0^{t+}\int_{|y|\geq 1}
      \Big(\int_0^1 A(X(s-)\!+\!\theta y)\!\cdot\! y\,d\theta\Big)\,
  N_X(dsdy) \nonumber\\
 &+&i\int_0^{t+}\int_{0<|y|<1}
       \Big(\int_0^1 A(X(s-)\!+\!\theta y)\!\cdot\! y\,d\theta\Big)\,
               \widetilde{N}_X(dsdy)\nonumber\\
 &+&i\int_0^{t} \!ds\, \hbox{\rm p.v.}\!\int_{0<|y|<1}
   \!\Big(\!\int_0^1\! A(X(s)\, +\!\theta y)\!\cdot\! y\,d\theta\Big)
          n(dy)\nonumber\\
 && \qquad\qquad\qquad\qquad \!+\!\!\int_0^t V(X(s))ds.\nonumber
\end{eqnarray}
\end{thm}

The proof of Theorem 3.2 is the same as that of Theorem 3.1.
We have only to replace
$A(X(s-)+\tfrac{y}{2})\!\cdot\! y$ by
$\displaystyle{\int_0^1} A(X(s-)+\theta y)\!\cdot\! y\,d\theta$
and consider
\begin{equation}
(F(t)g)(y) := \int_{{\bf R}^d} (e^{-t[\sqrt{-\Delta+m^2}-m]})(x-y)
      e^{[-i(y-x)\int_0^1 A((1-\theta)y+\theta x) d\theta -V(y)t]}g(y)dy,
\end{equation}
for which  note the second expression of the definition
(2.2) of $H_A^{(2)}$.  Etc.

%%%%%%%%%%%%%

\subsection{The case for $H^{(3)} := H^{(3)}_A + V$}%subsect3.3

The kinetic part $H^{(3)}_A$ is defined by operator-theoretical square root
of the Schr\"odinger operator
$S:=2H_A^{NR}+m^2,\,\, H_A^{NR}:= \tfrac12(-i\nabla-A(x))^2$.
We can say all information of $H^{(3)}_A$ is contained in $S:=2H_A^{NR}+m^2$
or the {\it nonrelativistic magnetic Schr\"odinger operator} $H_A^{NR}$.
So the problem is how to extract the information from it.
%\newline
For instance, the corresponding semigroup $e^{-t(H^{(3)}_A-m)}$ is completely
determined by $H^{NR}_A$ through theory of fractional powers
[Y, Chap.IX, 11, pp.259--261] as
\begin{eqnarray*}
e^{-t[H_A^{(3)}-m]}g &=&\left\{
\begin{array}{rl}
e^{mt} \int_0^{\infty} f_t(\lambda)
     e^{-\lambda[2H_A^{NR}+m^2]} g\,
    d\lambda, &\quad t>0,\\
 0, &\quad t=0
\end{array}\right. \\
f_t(\lambda) &=& \left\{
\begin{array}{rl}
  (2\pi i)^{-1} \int_{\sigma-i\infty}^{\sigma+i\infty}
    e^{z\lambda -tz^{1/2}}dz, &\quad \lambda \geq 0,\\
0, &\quad \lambda <0 \quad(\sigma >0).
\end{array}\right.
\end{eqnarray*}
Here $e^{-\lambda[2H_A^{NR}+m^2]}$ is represented
by the Feynman--Kac--It\^o formula, but we don't do it.

Instead, we note there is probabilistic counterpart of the above procedure
of going from Wiener process
($\equiv$nonrelativistic Schr\"odinger) to L\'evy process($\equiv$
(square root) relativistic Schr\"odinger).
It is {\it subordination} (by Bochner).

In this context, the problem of path integral for $e^{-t[H^{(3)}-m]}g$
was studied first by
DeAngelis, Serva and Rinaldi [AnSe-90], [AnRSe-91], then by [N-96, 97, 00]
with use of {\it subordination} of Brownian motion,
 and recently more extensively by
Hiroshima--Ichinose--L\H{o}rinczi [HILo-12, 13] (cf. [LoHB-11])
not only for magnetic relativistic Schr\"odinger operator
but also for {\it Bernstein functions} of magnetic
nonrelativistic Schr\"odinger operator even with spin.

Now, what is {\it subordination} ?

Start with the 1-dimensional {\it standard} Brownian motion
$B^1(t) \in C_0([0,\infty)\rightarrow {\bf R})$ with $B^1(0) =0$ and
$\mu_0$ the Wiener measure on $C_0([0,\infty)\rightarrow {\bf R})$
such that $e^{-t\tfrac12\xi^2}
= \displaystyle{\int_{C_0([0,\infty)\!\rightarrow\!{\bf R})}}
e^{iB^1(t)\xi} d\mu_0(B^1)$, then put
\begin{equation}%#(18)
 T(t) := \inf \{\, s > 0\,;\,
     B^1(s)+ m\,s = \,t\}, \qquad t\geq 0.
\end{equation}
Then $T(t)$ becomes a monotone, non-decreasing function on $[0,\infty)$
with $T(0)=0$, belonging to $D_0([0,\infty)\!\rightarrow\! {\bf R})$,
so that it is a 1-dimensional L\'evy process. This $T(t)$ is
what is called {\it subordinator}
([Sa-99, Chap.6, p.197], cf. [Sa-90]; [Ap-09, 1.3.2, p.52]),
which gives {\it time change}.
Let $\nu_0$ be the probability measure of the associated
process on space $D_0([0,\infty)\!\rightarrow\! {\bf R})$.

\bigskip\noindent
{\bf Lemma B}. (e.g. [Ap-09, p.54, Example 1.3.21, p.54, and
Exercise 2.1.10, p.96; cf. Theorem 2.2.9, p.95])
\begin{equation} %(3.16)
e^{-t[\sqrt{2\sigma+m^2}-m]}
= \int_{D_0([0,\infty)\rightarrow{\bf R})}e^{-T(t)\sigma} { d\nu_0(T)},
\qquad \sigma \geq 0
\end{equation}

We are in a position to give a path integral representation for
$e^{-t[H^{(3)}-m]}g$.

%\bigskip\noindent
\begin{thm} \mbox{\rm ([AnSe-90], [AnRSe-91], [N-96, 97, 00]; [HILo-12])}.%Thm3.3
\begin{eqnarray}%(3.17)
\quad({e^{-t[H^{(3)}-m]}g})(x) &\!=&\!
 \int\!\int_{\stackrel{\scriptstyle C_x([0,\infty)\rightarrow {\bf R}^d)}
   {\times D_0([0,\infty)\rightarrow{\bf R})}} e^{-{ S^{(3)}(t,B,T)}}
    \!g(B({ T(t)})) { d\mu_x(B) d\nu_0(T)},\\
{S^{(3)}(t,B,T)} &\!=& i\int_0^{{ T(t)}} \!\!A(B(s))\,dB(s)
       +\tfrac{i}{2}\int_0^{{ T(t)}}\!\! \hbox{\rm div} A(B(s))ds\nonumber\\
   &&\qquad\qquad\qquad\qquad\qquad\qquad\qquad
                             + \int_0^t \!\!V(B({ T(s)}))ds,\nonumber\\
&\!\equiv&
    i\!\int_0^{{ T(t)}} A(B(s))\circ
     dB(s)+ \int_0^tV(B({ T(s)}))ds \nonumber
\end{eqnarray}
%\noindent
{\it Here $C_x([0,\infty)\rightarrow {\bf R}^d)$ is the set of
continuous paths (Brownian motions)
$B: [0,\infty)\rightarrow {\bf R}^d$ with $B(0)=x$, and
{ $\mu_x$} is the Wiener measure on} $C_x([0,\infty)\rightarrow {\bf R}^d)$:
$$%\begin{equation}%$$ %#(22)
\exp\Big[-t\tfrac{\xi^2}{2}\Big]=
\int_{C_x([0,\infty)\rightarrow {\bf R}^d)}e^{i(B(t)-x)\cdot \xi}
d\mu_x(B) \qquad (m>0)
$$%\end{equation}%$$
\end{thm}

\bigskip
Before going to proof of Theorem 3.3, recall the Feynman--Kac--It\^o formula
[e.g. S-05] for the magnetic nonrelativistic Schr\"odinger operator
$H^{NR} := H_A^{NR}+ V := \tfrac{1}{2} (-i\nabla -A(x))^2 +V(x)\, $:
\begin{eqnarray}  %(3.18)
&&(e^{-tH^{NR}}\!g)(x)\\
&\!=\!& \int_{C_x([0,\infty)\rightarrow {\bf R}^d)}
   \!\!e^{-[i\int_0^t A(B(s)) dB(s)
      +\tfrac{i}{2} \int_0^t \hbox{\footnotesize div} A(B(s)) ds
      +\int_0^tV(B(s))ds]} \!g(B(t)) d\mu_x(B)\nonumber\\
&\!\equiv\!& \int_{C_x([0,\infty)\rightarrow {\bf R}^d)} \,
   \!e^{-[i\int_0^{t} A(B(s))\circ dB(s)+\int_0^tV(B(s))ds]}g(B(t))\, d\mu_x(B)
\,\,(\hbox{\rm Stratonovich})\nonumber
\end{eqnarray}

\bigskip\noindent
{\sl Proof of Theorem 3.3 (Sketch)}.
We use Lemma B, the spectral theorem for selfadjoint operator and
Feynman--Kac--It\^o formula above.
Note that $H_A^{(3)} = \sqrt{2H_A^{NR}+m^2}$.
 $\langle\cdot, \cdot \rangle$ stands the inner product of
Hilbert space $L^2({\bf R}^2)$. By spectral theorem for the selfadjoint operator
$H_A^{NR}$ (magnetic nonrelativistic Schr\"odinger operator with $V= 0$),
we have $H_A^{NR}
= \displaystyle{\int_{\hbox{\rm\scriptsize{Spec}}(H_A^{NR})}}
{ \sigma} \,dE({ \sigma})$.
Then for $f, g \in L^2({\bf R}^d)$
$$
\langle f, e^{-t  [H_A^{(3)}-m]}g \rangle =
\int_{\hbox{\rm \scriptsize{Spec}}(H_A^{NR})}
   e^{-t[\sqrt{2{ \sigma}+m^2}-m]} \,\langle f,
   dE({ \sigma})g \rangle
$$
By Lemma B and again by spectral theorem,
\begin{eqnarray*}
\langle f, e^{-t [H_A^{(3)}-m]}g \rangle
&=& \int_{\hbox{\rm \scriptsize{Spec}}(H_A^{NR})}
   \int_{D_0([0,\infty)\rightarrow {\bf R})}e^{-T(t){ \sigma}}
   d\nu_0(T) \, \langle f, dE({ \sigma})g \rangle\\
&=& \int_{D_0([0,\infty)\rightarrow {\bf R})}
      \langle f, e^{-T(t)H_A^{NR}} g \rangle\, d\nu_0(T).
\end{eqnarray*}
Applying { Feynman--Kac--It\^o} (with $V= 0$)
to $e^{-T(t) H_A^{NR}}g$ on the right-hand side,
\begin{eqnarray*}
\!\!\!\!&&\!\!\!\!\langle f,e^{-t  [H_A^{(3)}-m]}g\rangle \\
\!&\!=\!&\
  \int_{D_0([0,\infty)\rightarrow {\bf R})}\!\!\!\!\! d\nu_0(T)
 \!\!\int_{{\bf R}^d} \!\!\!dx \overline{f(\!B(0))}\!\!\!
 \int_{C_x([0,\infty)\rightarrow {\bf R}^d)}
 \!\!\!\!e^{-i\int_0^{T(t)}\!\!A(\!B(s))\circ dB(s)}\!g(\!B(T(t)))
  d\mu_x(\!B)\\
\!&\!=\!&\! \int_{{\bf R}^d} dx \overline{f(x)}
 \!\!\!\int_{D_0([0,\infty)\rightarrow {\bf R})}
 \!\!\int_{C_x([0,\infty)\rightarrow {\bf R}^d)}
   \!\!e^{-i\int_0^{T(t)} \!A(B(s))\circ dB(s)}\!g(\!B(\!T(t)))\, d\nu_0(T)
d\mu_x(B)
\end{eqnarray*}
Here note $B(0)=x$.
This proves the assertion when $V= 0$.

{ When $V\not= 0$}, with  partition of $[0,t]$:
$0=t_0<t_1< \cdots < t_n=t,\,\, t_j-t_{j-1}=t/n,$
we can express $e^{-t  [H^{(3)}-m]}g = e^{-t[(H_A^{(3)}-m)+V]}$
{ by Trotter--Kato formula or by Chernoff's theorem} with
$F(t) := e^{-t[H^{(3)}-m]}e^{-tV}$,
$$
e^{-t  [H^{(3)}-m]}g = \lim_{n\rightarrow \infty}
\big(e^{-(t/n)[H_A^{(3)}-m]}e^{-(t/n)V}\big)^ng,
$$
where convergence on the right-hand side is in strong sense.
Rewrite these $n$ operators product
by path integral on probability product measure
$\nu_0(T) \cdot\mu_x(B)$, then  we have (recall
$T(0)\!=\!T(t_0)\!=\!0,\,  B(0)\!=\!B(T(t_0))\!=\!x$),
\begin{eqnarray*}
&&\!\!\!\!\!
\langle f, \big(e^{-(t/n)[H_A^{(3)}-m]}e^{-(t/n)V}\big)^ng\rangle\\
&\!=\!& \int_{{\bf R}^d} dx
\int_{D_0([0,\infty)\rightarrow {\bf R})}\!\! d\nu_0(T)
\int_{C_x([0,\infty)\rightarrow {\bf R}^d)} \overline{f(B(0))}\\
&&\qquad\qquad\qquad
\times e^{-i\sum_{j=1}^n \int_{T(t_{j-1})}^{T(t_j)} A(B(s)\circ dB(s)}
e^{-\sum_{j=1}^n  V(B(T(t_{j}))\tfrac{t}{n}} g(B(t_n))\,d\mu_x(B) \\
\end{eqnarray*}
We see, as $n\rightarrow \infty$, that LHS converges to
$\langle f, e^{-t  [H_A^{(3)}-m]}g\rangle$, and
the right-hand side also converges to the goal formula as integral by
the product measure $dx \cdot \nu_0(T) \cdot \mu_x(B)$,
through Lebesgue theorem. This shows the weak convergence.
The strong convergence will also be shown.\qed

%%%%%%%%%%%%%%%%%%%%%%%%%%%%%%%%%%%
\subsection{Summary of three path intergal formulas} %sect3.4

Finally, as summary, we will collect the three path integral representation
formulas in Theorems 3.1, 3.2, 3.3, below, so as to be able to easily see
$x$-dependence. To do so, make change of space, probability measure and paths
by translation:

%%%%%%%%%%%
\smallskip
$D_x \!\!\rightarrow\!\! D_0$, $\lambda_x\!\!\rightarrow\!\! \lambda_0$,
$X(s)\!\! \rightarrow\!\! X(s)\!+\!x$, $B(s)\!\! \rightarrow\!\! B(s)\!+\!x$,
$B(T(s))\!\! \rightarrow\!\! B(T(s))\!+\!x$, then

\medskip
{\footnotesize
\begin{eqnarray*}
\hbox{(3.5)}: (e^{-t[H^{(1)}-m]}g)(x) &=&
 \int_{D_0([0,\infty)\rightarrow {\bf R}^d)} e^{-S^{(1)}(t,X)}
  g(X(t)+x)\,d\lambda_0 (X),\\ %#(11)
S^{(1)}(t,X)&=&i\int_0^{t+}\int_{|y|\geq 1}
                 A(X(s-)+x+\tfrac{y}{2})\!\cdot\! y\, N_X(dsdy) \nonumber\\
 &&    +i\int_0^{t+}\int_{0<|y|<1}
        A(X(s-)+x+\tfrac{y}{2})\!\cdot\! y\, \widetilde{N}_X(dsdy)\nonumber\\
 && +i\int_0^{t} ds\, \hbox{\rm p.v.}\! \int_{0<|y|<1}
        A(X(s)+x+\tfrac{y}{2})\!\cdot\! y\, n(dy) +\int_0^t V(X(s)+x)ds\,;\\
\hbox{(3.12)}: (e^{-t[H^{(2)}-m]}g)(x) &=&
  \int_{D_0([0,\infty)\rightarrow {\bf R}^d)} e^{-S^{(2)}(t,X)}
        g(X(t)+x)\,d\lambda_0 (X),\\
S^{(2)}(t,X) &=& i\!\int_0^{t+}\int_{|y|\geq 1}
\Big(\int_0^1 A(X(s-)\!+\!x\!+\!\theta y)\!\cdot\! y\,d\theta\Big)\, N_X(dsdy)
 \nonumber\\
 &&\!\! +i\!\int_0^{t+}\int_{0<|y|<1}
  \Big(\int_0^1 A(X(s-)\!+\!x\!+\!\theta y)\!\cdot\! y\,d\theta\Big)\,
               \widetilde{N}_X(dsdy)\nonumber\\
 &&\!\!
 +i\!\int_0^{t} \!ds\, \operatorname{{\scriptsize p.\!v.}}\!\!\int_{0<|y|<1}
 \!\!\Big(\!\int_0^1\!\! A(X(s)\!+\!x\!+\!\theta y)\!\cdot\!y \,d\theta\Big)
          n(dy) \!+\!\!\int_0^t \!\!V(X(s)\!+\!x)ds\,; \nonumber\\ %#(16)
\hbox{(3.17)}: (e^{-t[H^{(3)}-m]}g)(x)&=&
 \int\int_{\stackrel{\scriptstyle C_0([0,\infty)\rightarrow {\bf R}^d)}
{\times D_0([0,\infty)\rightarrow{\bf R})}} \, e^{-S^{(3)}(t,B,T)}
    g(B(T(t))\!+\!x)\, d\mu_0(B) d\nu_0(T),\\
S^{(3)}(t,B,T) &=&
 i\int_0^{T(t)} \!\!\!\!\!A(B(s)\!+\!x)\!\cdot \!dB(s)
   \!+\!\tfrac{i}{2}\!\int_0^{T(t)}\!\!\!\!\! \hbox{\rm div} A(B(s)\!+\!x)ds
        \!+\!\! \int_0^t \!\!V(B(T(s))\!+\!x)ds,\nonumber\\
 & \equiv&
  i\int_0^{T(t)} A(B(s)+x)\circ dB(s)+ \int_0^tV(B(T(s))+x)ds
\end{eqnarray*}
}

%%%%%%%%%%%%%%%%%%%%%%%%%%%%%
\subsection{Path integral formulas (3.5), (3.12) and (3.15)
as time-sliced approximation} %subsect3.5

In the proof of path integral formula (3.5) in Theorem 3.1,
we have used Chernoff's theorem to show Lemma A, i.e. that $F(t/n)^ng$,
the left-hand side of equality (3.10) converges to $e^{-tH^{(1)}}g$,
the left-hand side of (3.5).
We are now going to see how $F(t)$ in (3.6)/(3.11) comes out heuristically
for $L$ being the relativistic Schr\"odinger operators
$H^{(1)} = H_A^{(1)}+V$, $H^{(1)} = H_A^{(1)}+V$, but
there is a different situation for $H^{(3)} = H_A^{(3)}+V$.
For the details, we refer to [I-13, \S 4.2].

We use in what follows the time-sliced approximation, with partition
of the interval $[0,t]$ into $n$ small subintervals:
$0=t_0 <t_1 < \cdots <t_n =t$, only with {\sl equal width}
$t/n, \, t_j-t_{j-1}=t/n,\, 1\leq j \leq n$.

\medskip\noindent
{\bf (1)} {\sl For $L = H^{(1)} = H_A^{(1)}+V$.}
The second member of (3.10) can be heuristically rewritten
by the {\it imaginary-time phase space path integral} ([G-66], [M-78])
through time-sliced approximation, i.e. as the $n\!\rightarrow\! \infty$ limit
of the integral
\begin{eqnarray}
&&\displaystyle{\overbrace{\int_{{\bf R}^{2d}}\cdots\int_{{\bf R}^{2d}}}
^{\scriptsize{\mbox{$n$ times}}}}
e^{i\sum_{l=1}^n \big(X(t_l)-X(t_{l-1}))\big)\cdot\Xi(X(t_{l-1}))}\nonumber\\
 &&\qquad\qquad\qquad\qquad\times
   e^{-\tfrac{t}{n}\sum_{l=1}^n \big[\sqrt{\big(\Xi(t_{l-1})
       -A\big(\tfrac{X(t_{l-1})+X(t_{l})}2\big)\big)^2+m^2}-m
               +V(X(t_{l-1}))\big]} \nonumber\\
 &&\qquad\qquad\qquad\qquad\qquad\qquad\qquad\qquad\qquad\qquad\qquad
  \times g(X(0)) \prod_{j=1}^n \tfrac{d\Xi(t_{j-1})dX(t_{j-1})}{(2\pi)^d}
                       \nonumber\\
&=&\displaystyle{\overbrace{\int_{{\bf R}^{2d}}\cdots\int_{{\bf R}^{2d}}}
^{\scriptsize{\mbox{$n$ times}}}}
e^{i\sum_{l=1}^n (X(t_l)-X(t_{l-1}))\cdot\big(\Xi(t_{l-1})
         +A\big(\tfrac{X(t_{l-1})+X(t_{l})}2\big)\big)} \nonumber\\
 &&\qquad\qquad\quad\times
   e^{-\tfrac{t}{n}\sum_{l=1}^n \big[\sqrt{\Xi(t_{l-1})^2+m^2}-m
               +V(X(t_{l-1}))\big]} \nonumber%\\
 g(X(0)) \prod_{j=1}^n \tfrac{d\Xi(t_{j-1})dX(t_{j-1})}{(2\pi)^d}
                       \nonumber\\
&=&\displaystyle{\overbrace{\int_{{\bf R}^{2d}}\cdots\int_{{\bf R}^{2d}}}
^{\scriptsize{\mbox{$n$ times}}}}
e^{\sum_{l=1}^n \big\{i(x_l-x_{l-1})\cdot\big(\xi_{l-1}
                  +A\big(\tfrac{X(t_{l-1})+X(t_{l})}2\big)\big)
 -\tfrac{t}{n}[\sqrt{\xi_{l-1}^2+m^2}-m +V(x_{l-1})]\big\}} \nonumber\\
 &&\qquad\qquad\qquad\qquad\qquad\qquad\qquad\qquad\qquad\qquad\qquad
 \times g(x_0) \prod_{j=1}^n \tfrac{d\xi_{j-1}dx_{j-1}}{(2\pi)^d},
\end{eqnarray}%(3.19)
where, in the first equality, we made {\sl change of variables}:
$\Xi'(\cdot) := \Xi(\cdot) + A(X(\cdot),\, X'(\cdot) :=X(\cdot)$ on the
space of phase space paths, and then written $\Xi(\cdot), \,X(\cdot)$
again for $\Xi'(\cdot), \,X'(\cdot)$. In the second equality, we put
$\xi_j=\Xi(t_j),\, x_j = X({t_j}), \, j= 0,1, \dots, n-1$, and
$x=x_n =X(t_n)=X(t)$. Notice that here the assignment
$t_j \mapsto (\Xi(t_j), X(t_j))$ differs from the one used for (3.8).
 This is chronological, while that was anti-chronological.
Equation (3.19) is suggesting us how that functional of the path $X(\cdot)$,
 which is to be created as $e^{-S^{(1)}(t,x)}$ in (3.5) by
the approximation $F(t/n)^n$, does look.

Then the last member of (3.19) can be rewritten as
\begin{eqnarray}%
&&\displaystyle{\overbrace{\int_{{\bf R}^{2d}}\cdots\int_{{\bf R}^{2d}}}
^{\scriptsize{\mbox{$n$ times}}}}
\prod_{l=1}^n \Big\{\big[e^{i(x_l-x_{l-1})\cdot\xi_{l-1}}
e^{-\tfrac{t}{n}[\sqrt{\xi_{l-1}^2+m^2}-m]}\big] \nonumber\\
 &&\qquad\qquad\qquad\qquad
 \times e^{\big[iA\big(\tfrac{x_{l-1}+x_l}{2}\big)\cdot(x_l-x_{l-1})
           -V\big(\tfrac{x_{l-1}+x_l}{2}\big)\tfrac{t}{n}\big]}\Big\}
  g(x_0) \prod_{j=1}^n \tfrac{d\xi_{j-1}dx_{j-1}}{(2\pi)^d}\nonumber\\
&=&\displaystyle{\overbrace{\int_{{\bf R}^{d}}\cdots\int_{{\bf R}^{d}}}
^{\scriptsize{\mbox{$n$ times}}}}
\prod_{l=1}^n \Big\{\big[e^{-\tfrac{t}{n}[\sqrt{-\Delta+m^2}-m]}\big]
         (x_l-x_{l-1})\nonumber\\
  &&\qquad\qquad\qquad\qquad
\times e^{\big[iA\big(\tfrac{x_{l-1}+x_l}{2}\big)\cdot(x_l-x_{l-1})
           -V\big(\tfrac{x_{l-1}+x_l}{2}\big)\tfrac{t}{n}\big]}\Big\}
             g(x_0)\, dx_0 dx_1 \cdots dx_{n-1},
\end{eqnarray}%(3.20)
with $x=x_n$, where we have performed all the $d\xi_j$ integrations.
The result is nothing but $F(t/n)^n g$ in (3.10) with $F(t)$ in (3.7).

\medskip\noindent
{\bf (2)} {\sl For $L = H^{(2)} = H_A^{(2)}+V$.}
Similar treatment is valid for $L = H^{(2)} = H_A^{(2)}+V$, where
we may consider for $H^{(2)}$ with
$$
\int_0^1 A\big((1-\theta)X(t_{l-1})+ \theta X(t_l)\big)d\theta
$$
{in place of}
$$
A\big(\tfrac{X(t_{l-1})+X(t_l)}2\big)
$$
for $H^{(1)}$ on  each subinterval $[t_{j-1},t_j]$.
The same arguments as for $L = H^{(1)}$ above above will show
the expression (3.12) is also obtained heuristically
through time-sliced approximation with $F(t)$ in (3.14).

\medskip\noindent
{\bf (3)} {\sl For $H^{(3)} = H_A^{(3)}+V$.}
In this case, formula (3.17) does not seem to be one which
can be heuristically obtained, probably because $H_A^{(3)}$ cannot
be so explicitly well expressed by a pseudo-differential operator
defined through a certain
{\sl tractable symbol} as $H_A^{(1)}$ and $H_A^{(2)}$.

Indeed, for the semigroups $e^{-t[H_A^{(1)}+V]}$ and $e^{-t[H_A^{(2)}+V]}$,
take (3.7)/(3.11) as $F(t)$, we could show that
$F(t/n)^n \rightarrow e^{-t[H_A^{(j)}+V]}$ strongly for $j=1,2$.
But for the semigroup $e^{-t[H_A^{(3)}+V]}$, such an interpretation
does not seem possible.

%%%%%%%%%%%%%%%%%%%%%%%%%%%
%\newpage
\section{Some observation on
Chernoff's theorem and path integral by time-sliced approximation}
%sect4.

It is well-known that, for the solution of Schr\"odinger equation,
the Trotter--Kato product formula can simply and plainly
give a, though naive, meaning to its path integral
 representation by {\it time-sliced approximation}, if the Schr\"odinger operator
has no {\it magnetic vector potential} but only electric
scalar potential $V(x)$. However, if it has also {\it magnetic vector potential}
$A(x)$, it does not seem to go well, and then we need Chernoff's theorem.
Our aim is to observe how useful and effective a tool Chernoff's theorem
is to give a meaning to path integral formulas by
{\it time-sliced approximation}, guaranteeing its convergence. For this aspect,
we also refer to [BoBuScSm-11].

For our convenience, we begin this section with restating the Chernoff's theorem,
though already done in \S 3.1.
Notice that Trotter--Kato product formula follows from Chernoff's theorem, but
the converse is not valid.

\medskip\noindent
{\bf Chernoff's Theorem}. [Ch-74]
$\,\,\,${\it Let $F$ be a strongly continuous function on $[0,\infty)$
 with values in the Banach space ${\cal L}({\bf X})$
of bounded linear operators on a Banach space ${\bf X}$.
Assume that $F$ further satisfies the following conditions}:
(i) {\it $F(0) = I$ ($I$: identity operator on ${\bf X}$), and there exists
a real $a$ such that $\|F(t)\| \leq e^{at}$  for all $t\geq 0$;}

\noindent
(ii) {\it The linear operator $F'(0)\upharpoonright_{D[F'(0)]}$ is closable,
and the closure $\overline{F'(0)} := L$
      generates a strongly continuous semigroup $e^{-tL} $.}

{\it Then
$F(t/n)^n$ converges to $e^{-tL}$  strongly, as $n\rightarrow\infty$,
   uniformly on each finite interval in  $t \geq 0$. }

\bigskip
The content of this section is almost independent of the three
relativistic Schr\"odinger operators $H^{(1)}$, $H^{(2)}$ and $H^{(3)}$
and their path integral representation formulas, about which
we have already discussed enough up to the previous section \S3.
In this section, we will study further this wisdom with several
other evolution equations in quantum mechanics to watch their corresponding
path integral representation formulas.
We first treat the case of strong convergence and next the case of convergence
in norm and/or pointwise for the integral kernels.

Throughout this section again,
the time-sliced approximation, with partition of the interval $[0,t]$
into $n$ small subintervals:  $0=t_0 <t_1 < \cdots <t_n =t$, is used
only with {\sl equal width} $t/n, \, t_j-t_{j-1}=t/n,\, 1\leq j \leq n$.

%%%%%%%%%%%%%%%%%%%%%%%%%%%
\subsection{Time-sliced approximation in strong topology} %subsect4.1

We consider, first, the time-sliced approximation
for the solution of Schr\"odinger equation in real and/or imaginary time,
only with scalar potential, that is, {\sl without magnetic vector potential},
and see it strongly converge by Trotter--Kato product formula as well
as Chernoff's theorem.
Next, we come to consider the Schr\"odinger equation and Dirac equation
{\sl in presence of magnetic vector potential} and realize in turn to need
to use Chernoff's theorem.

\newpage
\subsubsection{Schr\"odinger operator with scalar potential $V(x)$}
%subsubsect4.1.1

The operator concerned is  $H_V := -\tfrac12\Delta+V$ in $L^2({\bf R}^3)$.
Put
\begin{eqnarray}%(4.1),(4.2)
&&(F(t)g)(x) := (e^{-it(-\tfrac12\Delta)}e^{-itV})(x)
  = \int [e^{-it(-\tfrac12\Delta)}](x-y)e^{-itV(y)}g(y) dy,\\
&&(G(t)g)(x) := (e^{-t(-\tfrac12\Delta)}e^{-tV}g)(x)
   = \int [e^{-t(-\tfrac12\Delta)}](x-y)e^{-tV(y)}g(y) dy,
\end{eqnarray}
where $[e^{-it(-\tfrac12\Delta)}](x-y)$ and
$[e^{-t(-\tfrac12\Delta)}](x-y)$ stand for the the integral kernels
of the Schr\"odinger unitary group $e^{-it(-\tfrac12\Delta)}$
and Schr\"odinger semigroup $e^{-t(-\tfrac12\Delta)}$,
respectively.

Under certain reasonable conditions on $V(x)$, it holds
in strong resolvent sense in $L^2({\bf R}^3)$ as $t\downarrow 0$ that
$\frac{I-F(t)}{t}$ converges to $iH_V$,
while $\frac{I-G(t)}{t}$ converges to $H_V$.
Then by Chernoff's
theorem or in this case by Trotter--Kato product formula,
we have, for $g \in L^2({\bf R}^3)$,
\begin{eqnarray}%(4.3),(4.4)
F(t/n)^n g \rightarrow &e^{-it[-\tfrac12\Delta+V]}g&\,,
                                \qquad{strongly},\\
G(t/n)^n g \rightarrow &e^{-t[-\tfrac12\Delta+V]}g&\,,
                                \qquad{strongly}\,,
\end{eqnarray}
as $n\rightarrow \infty$.
On the other hand, $e^{-it[-\tfrac12\Delta+V]}g$
should be given by the {\it configuration space path integral}
through time-sliced approximation as the $n\!\rightarrow\! \infty$ limit of
the integral
\begin{eqnarray}%(4.5)
&& C_n
 \displaystyle{\overbrace{\int_{{\bf R}^{d}}\cdots\int_{{\bf R}^{d}}}
^{\scriptsize{\mbox{$n$ times}}}}
   e^{i\sum_{l=1}^n \big[\tfrac12\big(\tfrac{X(t_l)-X(t_{l-1})}{t/n}\big)^2
                        -V(X(t_{l-1}))\big]\tfrac{t}{n}}
  g(X(0)) \prod_{j=1}^{n} d(X(t_{j-1}))\\
&=& C_n
 \displaystyle{\overbrace{\int_{{\bf R}^{d}}\cdots\int_{{\bf R}^{d}}}
^{\scriptsize{\mbox{$n$ times}}}}
\prod_{j=1}^n \big[e^{i\tfrac{t}{n}\tfrac12\big(\tfrac{x_j-x_{j-1}}{t/n}\big)^2}
 e^{-i\tfrac{t}{n}V(x_{j-1})}\big]g(x_0) dx_0dx_1 \cdots dx_{n-1},\nonumber
\end{eqnarray}
with some renormalization constant $C_n$ depending on $t$,
where we put $x_j = X(t_j), \, j= 0,1, \dots, n-1$, and $x=x_n =X(t_n)=X(t)$.
Taking $C_n = \big(\tfrac{i}{2\pi t/n}\big)^{3n/2}$, this is
what is meant by $F(t/n)^ng$.

Similarly, $e^{-t[-\tfrac12\Delta+V]}g$
should be given by the {\it configuration space imaginary-time path integral}
through time-sliced approximation as the $n\!\rightarrow\! \infty$ limit of
the integral
\begin{eqnarray}%(4.6)
&&\ C'_n
 \displaystyle{\overbrace{\int_{{\bf R}^d}\cdots\int_{{\bf R}^d}}
^{\scriptsize{\mbox{$n$ times}}}}
e^{-\sum_{l=1}^n \big[\tfrac12\big(\tfrac{X(t_l)-X(t_{l-1})}{t/n}\big)^2
                        +V(X(t_{l-1}))\big]\tfrac{t}{n}}g(X(0))
  \prod_{j=1}^{n} d(X(t_{j-1}))\\
&=& C'_n
 \displaystyle{\overbrace{\int_{{\bf R}^d}\cdots\int_{{\bf R}^d}}
^{\scriptsize{\mbox{$n$ times}}}}
\prod_{j=1}^n \big[e^{-\tfrac{t}{n}\tfrac12\big(\tfrac{x_j-x_{j-1}}{t/n}\big)^2}
   e^{-\tfrac{t}{n}V(x_{j-1})}\big]g(x_0) dx_0dx_1 \cdots dx_{n-1},\nonumber
\end{eqnarray}
with some renormalization constant $C'_n$ depending on $t$,
where put $x_j = X(t_j), \, j= 0,1, \dots, n-1$, and $x=x_n =X(t_n)=X(t)$.
Taking $C'_n = \big(\tfrac{1}{2\pi t/n}\big)^{3n/2}$,
this is what is meant by
$G(t/n)^ng$.

%%%%%%%%%%%%%%%%%%%%
%\newpage
\subsubsection{Schr\"odinger operator with
vector and scalar potentials $A(x)$ and $V(x)$}%subsubsect4.1.2

The operator concerned is $H_{A,V} := \tfrac12(-i\nabla-A(x))^2+V$ in
$L^2({\bf R}^3)$.
Put
\begin{eqnarray}%(4.7).(4.8)
&&(F(t)g)(x) := \int [e^{-it\tfrac12(-\Delta)}](x-y)e^{i[A(\tfrac{x+y}2)(x-y)
               -V(\tfrac{x+y}2)t]} g(y)dy\, ;\\
&&(G(t)g)(x) := \int [e^{-t\tfrac12(-\Delta)}](x-y)e^{iA(\tfrac{x+y}2)(x-y)
               -V(\tfrac{x+y}2)t} g(y)dy\,.
\end{eqnarray}

%%%%%%%%%
%\newpage
\noindent
Then, under certain reasonable conditions on $A(x)$ and $V(x)$,
though one cannot use Trotter--Kato product formula because of presence
of the vector potential $A(x)$, we have by Chernoff's theorem instead that 
as $n\rightarrow \infty$,
\begin{eqnarray}%(4.9),(4.10)
F(t/n)^n g \rightarrow &e^{-it[\tfrac12(-i\nabla-A(x))^2+V]}g&\,,
                                       \qquad{strongly},\\
G(t/n)^n g \rightarrow &e^{-t[\tfrac12(-i\nabla-A(x))^2+V]}g&\,,
                                       \qquad{strongly}.
\end{eqnarray}
On the other hand, $e^{-it[-\tfrac12(-i\nabla-A(x))^2+V]}g$
should be given by the {\it phase space path integral} ([G-66], [M-78])
through time-sliced approximation. We make the same argument
for $H_{A,V}$ as used in (3.19) through (3.20) for the relativistic
Schr\"odinger operator $H^{(1)} = H_A^{(1)}+V$, but
{\sl here (and also below in \S4.2.3), for simplicity,
by skipping the step of performing the change of variables
(on the space of phase space paths) inside (3.19).}
Then $e^{-it[-\tfrac12(-i\nabla-A(x))^2+V]}g$
should be reached as the $n\!\rightarrow\! \infty$ limit of the integral
\begin{eqnarray}%(4.11)
&&\quad
\displaystyle{\overbrace{\int_{{\bf R}^{2d}}\!\cdots\!\int_{{\bf R}^{2d}}}
^{\scriptsize{\mbox{$n$ times}}}} e^{i\sum_{l=1}^n
 \big[(X(t_l)-X(t_{l-1}))\cdot\Xi(t_{l-1})
                     -\tfrac{t}{n}\tfrac{\Xi(t_{l-1})^2}2\big]}\\
  &&\times e^{i\sum_{l=1}^n
   \big[A\big(\tfrac{X(t_l)+X(t_{l-1})}{2}\big)\cdot(X(t_l)-X(t_{l-1}))
    -\tfrac{t}{n}V(X(t_{l-1}))\big]} g(X(0))
   \prod_{j=1}^{n} \tfrac{d\Xi(t_{j-1})dX(t_{j-1})}{(2\pi)^3}\nonumber\\
&=&\displaystyle{\overbrace{\int_{{\bf R}^{2d}}\!\cdots\!\int_{{\bf R}^{2d}}}
 ^{\scriptsize{\mbox{$n$ times}}}}
 \prod_{j=1}^n \big\{e^{\big[i(x_j-x_{j-1})\cdot\xi_{j-1}
    -i\tfrac{t}{n}\tfrac{\xi_{j-1}^2}2\big]}
    e^{i\big[A\big(\tfrac{x_j+x_{j-1}}{2}\big)\cdot(x_j-x_{j-1})
    -\tfrac{t}{n}V(x_{j-1})\big]}\big\} g(x_0) \nonumber\\
 &&\qquad\qquad\qquad\qquad\qquad\qquad\qquad\qquad\qquad\qquad\qquad
   \times \tfrac{d\xi_0dx_0}{(2\pi)^3}\tfrac{d\xi_1dx_1}{(2\pi)^3}
   \cdots \tfrac{d\xi_{n-1}dx_{n-1}}{(2\pi)^3} \nonumber\\
&=&\displaystyle{\overbrace{\int_{{\bf R}^{d}}\!\cdots\!\int_{{\bf R}^{d}}}
 ^{\scriptsize{\mbox{$n$ times}}}}
 \prod_{l=1}^n \big\{[e^{-i\tfrac{t}{n}\tfrac12(-\Delta)](x_l-x_{l-1})}
    e^{i\big[A\big(\tfrac{x_l+x_{l-1}}{2}\big)\cdot(x_l-x_{l-1})
    -\tfrac{t}{n}V(x_{l-1})\big]}\big\} g(x_0) \nonumber\\
 &&\qquad\qquad\qquad\qquad\qquad\qquad\qquad\qquad\qquad\qquad\qquad
   \times dx_0 dx_1\cdots dx_{n-1}, \nonumber
\end{eqnarray}
where put $\xi_j=\Xi(t_j),\, x_j = X({t_j}), \, j= 0,1, \dots, n-1$,
and $x=x_n =X(t_n)=X(t)$.
This is what is meant by $G(t/n)^ng$.

Similarly, $e^{-t[-\tfrac12(-i\nabla-A(x))^2+V]}g$
should be given by the {\it imaginary-time phase space path integral}
through time-sliced approximation as the $n\!\rightarrow\! \infty$ limit of
the integral
\begin{eqnarray}%(4.12)
&&\quad
\displaystyle{\overbrace{\int_{{\bf R}^{2d}}\!\cdots\!\int_{{\bf R}^{2d}}}
^{\scriptsize{\mbox{$n$ times}}}} e^{\sum_{l=1}^n
 \big[i(X(t_l)-X(t_{l-1}))\cdot\Xi(t_{l-1})
                       -\tfrac{t}{n}\tfrac12\Xi(t_{l-1})^2\big]}\\
  &&\times e^{\sum_{l=1}^n
   \big[iA\big(\tfrac{X(t_l)+X(t_{l-1})}{2}\big)\cdot(X(t_l)-X(t_{l-1}))
    -\tfrac{t}{n}V(X(t_{l-1}))\big]} g(X(0))
  \prod_{j=1}^{n} \tfrac{d\Xi(t_{j-1})dX(t_{j-1})}{(2\pi)^3}\nonumber\\
&=&\displaystyle{\overbrace{\int_{{\bf R}^{2d}}\!\cdots\!\int_{{\bf R}^{2d}}}
 ^{\scriptsize{\mbox{$n$ times}}}}
 \prod_{j=1}^n \big\{e^{[i(x_j-x_{j-1})\cdot\xi_{j-1}
                             -\tfrac{t}{n}\tfrac12\xi_{j-1}^2]}
    e^{\big[iA\big(\tfrac{x_j+x_{j-1}}{2}\big)\cdot(x_j-x_{j-1})
    -\tfrac{t}{n}V(x_{j-1})\big]}\big\} g(x_0) \nonumber\\
 &&\qquad\qquad\qquad\qquad\qquad\qquad\qquad\qquad\qquad\qquad\qquad
   \times \tfrac{d\xi_0dx_0}{(2\pi)^3}\tfrac{d\xi_1dx_1}{(2\pi)^3}
   \cdots \tfrac{d\xi_{n-1}dx_{n-1}}{(2\pi)^3} \nonumber\\
&=&\displaystyle{\overbrace{\int_{{\bf R}^{d}}\!\cdots\!\int_{{\bf R}^{d}}}
 ^{\scriptsize{\mbox{$n$ times}}}}
 \prod_{j=1}^n \big\{[e^{-\tfrac{t}{n}\tfrac{\Delta}2}](x_j-x_{j-1})
    e^{\big[iA\big(\tfrac{x_j+x_{j-1}}{2}\big)\cdot(x_j-x_{j-1})
    -\tfrac{t}{n}V(x_{j-1})\big]}\big\} g(x_0) \nonumber\\
 &&\qquad\qquad\qquad\qquad\qquad\qquad\qquad\qquad\qquad\qquad\qquad
   \times  dx_0 dx_1\cdots dx_{n-1}, \nonumber
\end{eqnarray}
where put $\xi_j=\Xi(t_j),\, x_j = X({t_j}), \, j= 0,1, \dots, n-1$,
and $x=x_n =X(t_n)=X(t)$.
This is what is meant by $G(t/n)^ng$.
Equation (4.12) is suggesting us how that functional of the (Brownian) path
$B(\cdot)$, which is to be created as the integrand of the
Feynman--Kac--It\^o formula (3.18) by the approximation $G(t/n)^n$,
does look (See [S-05, (15.1-2), p.159]).

%%%%%%%%%%%%%%%%%
%\bigskip
\subsubsection{Dirac operator with vector and scalar potentials
$A(x)$ and $V(x)$}%subsubsect4.1.3

The operator concerned is $\alpha\cdot(-i\nabla-A) +m\beta +V$
 in $L^2({\bf R}^3;{\bf C}^4)$ where $\alpha := (\alpha_1,\alpha_2,\alpha_3)$
and $\beta$ are Dirac four matrices.
Put 
\begin{eqnarray}%(4.13)
(F(t)f)(x) 
:&=& \int_{{\bf R}^3} K^{\operatorname{\tiny{Dirac}}}(t,x-y)
    e^{i[A(\tfrac{x+y}2)(x-y)-V(\tfrac{x+y}2)t]}f(y)dy \nonumber\\
 &=& \int_{{\bf R}^3} [e^{-it(\alpha\cdot(-i\nabla)+m\beta)}](x-y)
    e^{i[A(\tfrac{x+y}2)(x-y)-V(\tfrac{x+y}2)t]}f(y)dy
\end{eqnarray}
for $f \in L^2({\bf R}^3;{\bf C}^4)$,
where $K^{\operatorname{\tiny{Dirac}}}(t,x-y) 
        :=[e^{-it(\alpha\cdot(-i\nabla)+m\beta)}](x-y)$ 
is the integral kernel of the unitary group of {\it free} Dirac operator
$\alpha\cdot(-i\nabla)+m\beta$.
Then, under certain reasonable conditions on $A(x)$ and $V(x)$,
we have by Chernoff's theorem that as $n\rightarrow \infty$,
\begin{equation}%(4.14)
F(t/n)^nf \rightarrow e^{-it[(\alpha\cdot(-i\nabla-A)+m\beta)+V]}f\,,
                                           \qquad{strongly}.
\end{equation}
On the other hand, $e^{-it[\alpha\cdot(-i\nabla-A) +m\beta +V]}f$
should be given by the {\it phase space path integral}
through time-sliced approximation as the $n\!\rightarrow\! \infty$ limit
of the integral
\begin{eqnarray}%(4.15)
&&\qquad
\displaystyle{\overbrace{\int_{{\bf R}^{6}}\!\cdots\!\int_{{\bf R}^{6}}}
^{\scriptsize{\mbox{$n$ times}}}} e^{i\sum_{l=1}^n
 \big[(X(t_l)-X(t_{l-1}))\cdot\Xi(t_{l-1})
                -\tfrac{t}{n}(\alpha\cdot\Xi(t_{l-1}) +m\beta)\big]}\\
  &&\times e^{i\sum_{l=1}^n
   \big[A\big(\tfrac{X(t_l)+X(t_{j-1})}{2}\big)\cdot(X(t_l)-X(t_{l-1}))
    -\tfrac{t}{n}V(X(t_{l-1}))\big]} f(X(0))
   \prod_{j=1}^{n} \tfrac{d\Xi(t_{j-1})dX(t_{j-1})}{(2\pi)^3}\nonumber\\
&=&\displaystyle{\overbrace{\int_{{\bf R}^{6}}\!\cdots\!\int_{{\bf R}^{6}}}
 ^{\scriptsize{\mbox{$n$ times}}}}
 \!\prod_{j=1}^n\! \big\{e^{\big[i(x_j-x_{j-1})\cdot\xi_{j-1}
    -i\tfrac{t}{n}(\alpha\cdot\xi_{j-1}+m\beta)\big]}
    e^{i\big[A\big(\tfrac{x_j+x_{j-1}}{2}\big)\cdot(x_j-x_{j-1})
    -\tfrac{t}{n}V(x_{j-1})\big]}\big\} \nonumber\\
 &&\qquad\qquad\qquad\qquad\qquad\qquad\qquad\qquad\qquad\qquad\,\,\,
   \times f(x_0)\tfrac{d\xi_0dx_0}{(2\pi)^3}\tfrac{d\xi_1dx_1}{(2\pi)^3}
   \cdots \tfrac{d\xi_{n-1}dx_{n-1}}{(2\pi)^3} \nonumber\\
&=&\displaystyle{\overbrace{\int_{{\bf R}^{3}}\!\cdots\!\int_{{\bf R}^{3}}}
 ^{\scriptsize{\mbox{$n$ times}}}}
 \prod_{j=1}^n \big\{
    \big[e^{-i\tfrac{t}{n}(\alpha\cdot(-\nabla)+m\beta)}\big](x_j-x_{j-1})
    \big[e^{i\big[A\big(\tfrac{x_j+x_{j-1}}{2}\big)\cdot(x_j-x_{j-1})
    -\tfrac{t}{n}V(x_{j-1})\big]}\big\} \nonumber\\
 &&\qquad\qquad\qquad\qquad\qquad\qquad\qquad\qquad\qquad\qquad\qquad\qquad
   \times f(x_0) dx_0 dx_1\cdots dx_{n-1} \nonumber\\
&=&\displaystyle{\overbrace{\int_{{\bf R}^{3}}\!\cdots\!\int_{{\bf R}^{3}}}
 ^{\scriptsize{\mbox{$n$ times}}}}
 K^{\operatorname{\tiny{Dirac}}}(\tfrac{t}{n},x_n-x_{n-1}) 
 K^{\operatorname{\tiny{Dirac}}}(\tfrac{t}{n},x_{n-1}-x_{n-2})
 \cdot\,\cdots\,\cdot  
 K^{\operatorname{\tiny{Dirac}}}(\tfrac{t}{n},x_1-x_0) \nonumber\\
 &&\qquad\qquad\qquad \times 
 \big\{e^{i\sum_{j=1}^n \big[A\big(\tfrac{x_j+x_{j-1}}{2}\big)\cdot(x_j-x_{j-1})
    -\tfrac{t}{n}V(x_{j-1})\big]}\big\} 
     \,f(x_0)\, dx_0 dx_1\cdots dx_{n-1}. \nonumber
\end{eqnarray}
Here we have put $\xi_j=\Xi(t_j),\, x_j = X({t_j}), \, j=  0,1, \dots, n-1$,
$x=x_n =X(t_n)=X(t)$, and in the last equality, we have performed
all the $d\xi_j$ integrations. The last member of (4.15) is nothing but what 
is meant by $F(t/n)^nf$, and is suggesting us what a kind of functional of 
the path $X(\cdot)$ the expected path integral formula should turn out to have in 
its integrand. For instance, since $n\rightarrow \infty$,
$$\sum_{j=1}^n \big[A\big(\tfrac{x_j+x_{j-1}}{2}\big)\cdot(x_j-x_{j-1})
                   -\tfrac{t}{n}V(x_{j-1})\big]
 \rightarrow \int_0^t [A(X(s))\cdot dX(s) - V(X(s))ds],
$$
we should have an expression
\begin{eqnarray*}
&&\big\langle f_1, 
        e^{-it[\alpha\cdot(-i\nabla-A) +m\beta +V]} f_2 \big\rangle\\
&=& \int\int_{{\Bbb R}^3\times{\Bbb R}^3} dxdy\, 
\big\langle f_2(x),\, 
d\nu^{\operatorname{\tiny{Dirac}}}_{t,x;0,y}(X)\, 
     e^{i\int_0^t [A(X(s))\cdot dX(s)- V(X(s))ds]}f_1(y)\big\rangle,
\end{eqnarray*}
for all functions $f_1,\, f_2$, say, in the Schwartz space 
${\cal S}({\Bbb R}^3; {\Bbb C}^4)$, 
if there should exist a $3\times 3$-matrix-valued (countable additive) measure 
$\nu^{\operatorname{\tiny{Dirac}}}_{t,x;0,y}(X)$ on the space of 
Lipschitz-continuous paths $[0,t]\ni s \mapsto X(s) \in {\Bbb R}^3$ 
with $X(0)=y,\, X(t)=x$. However, no such measure
$\nu^{\operatorname{\tiny{Dirac}}}_{t,x;0,y}(X)$ can exist for this 3-dimensional 
Dirac operator $\alpha\cdot(-i\nabla-A) +m\beta +V$, 
although it can for 1-dimensional Dirac operator 
instead (cf. [I-82, 84], [ITa-84, 87], [I-93]). 

\subsection{Time-sliced approximation in norm and pointwise}%4.2

%\bigskip
\subsubsection{Trotter--Kato product formula and Chernoff's theorem
in norm}%subsect4.2.1

In [IT-01, ITTZ-01], we proved the selfadjoint Trotter--Kato product
formula {\it in norm}, i.e. {\it in operator norm}:

{\it If $A$ and $B$ are nonnegative selfadjoint operators
in a Hilbert space such that their operator sum $C:=A+B$ is also selfadjoint
with domain $D[C] := D[A]\cap D[B]$, then as $n\rightarrow \infty$,
$(e^{-\tfrac{t}{n}A}e^{-\tfrac{t}{n}B})^n$
as well as
$(e^{-\tfrac{t}{2n}B}e^{-\tfrac{t}{n}A}e^{-\tfrac{t}{2n}B})^n$
converges to $e^{-tC}$  in operator norm,
with optimal error estimate $O(n^{-1})$.
This means nothing but that $F(t/n)^n \rightarrow e^{-tC}$ in operator norm,
with $F(t) := e^{-tA}e^{-tB}$ or $F(t) := e^{-tB/2}e^{-tA}e^{-tB/2}$.}

\smallskip
Applying this result to the Schr\"odinger semigroup with
$H_V := -\frac12\Delta +V$, where $V(x) \geq 0$ and $H_V$ becomes
a selfadjoint operator in $L^2({\bf R}^d)$ with domain
$D[H_V] = D[\Delta]\cap D[V]$, we have
\begin{eqnarray*}
(e^{-\tfrac{t}{n}\tfrac12(-\Delta)}e^{-\tfrac{t}{n}V})^n
\,\, &\rightarrow& \,\,e^{-tH_V}, \qquad{in\,\, operator\,\, norm}\,,\\
(e^{-\tfrac{t}{2n}V}e^{-\tfrac{t}{n}\tfrac12(-\Delta)}
    e^{-\tfrac{t}{2n}V})^n \,\, &\rightarrow& \,\,e^{-tH_V},
                                  \qquad{in\,\, operator\,\, norm}\,,
\end{eqnarray*}
as $n\rightarrow \infty$, with error estimate $O(n^{-1})$.

\smallskip
The proof of this operator-norm version of Trotter--Kato product formula
is thanks to an operator-norm version of Chernoff's theorem,
{\sl even} with error estimate, established also in [IT-01] (cf. [NeZ-99]).
Only part of it without error estimate is given here.

\smallskip\noindent
{\bf Chernoff's Theorem in operator norm}. %([IT-01], c.f. [NeZ-99])
$\,\,\,${\it Let $\{F (t)\}_{t \geq0}$ be a family of selfadjoint
operators in a Hilbert space with $0 \leq F (t) \leq 1$.  Then if
\begin{align*}
&\|\big(1 + t^{-1}(I-F(t))\big)^{-1} - (1 + C)^{-1}\| \rightarrow 0,
      \quad t \downarrow 0,\\
\intertext{with $C$ some nonnegative selfadjoint operator, then}
&\|F (t/n)^n - e^{-tC}\| \rightarrow 0, \quad n \rightarrow \infty.
\end{align*}
}

As for the {\it unitary Trotter product formula in operator norm},
it does not in general hold. For some counterexamples, see [I-03, pp.88--90].
 However, there are some special cases where it holds for
the unitary groups for the Dirac operator and the relativistic Schr\"odinger
operator with {\it suitable potentials}. For the details, see [IT-04a].

%\bigskip
\subsubsection{Time-sliced approximation for Schr\"odinger equation in real
and imaginary time  --- convergence in norm and pointwise}%subsubsect4.2.2

As touched on only briefly at the end of \S4.2.1 just above,
the unitary Trotter product formula in norm, i.e. in operator norm
does not hold for the nonrelativistic Schr\"odinger operator
$H_V = -\frac12\Delta+V$ considered in \S4.1.1.

However, we want to discuss a little more how about the convergence
in operator norm and/or pointwise for the integral kernels by time-sliced
approximation and to observe some remarkable fact on the error estimate
of this approximation comparing the cases for the {\it real-time}
and {\it imaginary-time} nonrelativistic Schr\"odinger equations.

First, for the {\it real-time} nonrelativistic Schr\"odinger equation
$i\frac{\partial}{\partial t} \psi(t,x) = H_V \psi(t,x)$,
we visit Fujiwara's result [Fu-79, 80], in particular, book
[Fu-99, Theorems 4.22, 4.26, 5.4.1 (pp.79, 82, 105)]
or survey [Fu-12, Theorems 3.3, 3.4 (p.105)],
[FuKu-06, Theorem 2(p.843)] (cf. [Ku-04], [FuKu-05]). He made use of
a sophisticated way of time-sliced approximation for Feynman path integral
to construct the fundamental solution $e^{-itH_V}(x,y)$,  i.e.
the integral kernel of the Schr\"odinger unitary group $e^{-itH_V}$.
It is a much more elaborate time-sliced approximation than the one
naturally stemming from  the Trotter product formula.

For explanation, let $V(x)$ be a smooth function
satisfying $|\partial^{\alpha} V(x)|  \leq
C_{\alpha }(1+x^{2})^{(2 - |\alpha|)_+/2}$ for every multi-index
$\alpha $ with constant $C_{\alpha}$,
though $V(x)$ need not be bounded below. [For instance, this condition
is satisfied by $V(x) = \pm |x|^2$.]
Put
\begin{equation}%(4.16)
 (E(t)\varphi)(x) = (2 \pi it)^{-d/2} \int_{{\bf R}^d}
    e^{iS(t,x,y)} \varphi(y) dy
\end{equation}
for $\varphi \in C_0^{\infty}({\bf R}^d)$,  with action
$S(t,x,y) =
\int_0^t [\tfrac{1}{2} (d\overline{X}(s)/ds)^2 - V(\overline{X}(s))] ds$,
where $\overline{X}(s)$ is the classical trajectory starting
at $\overline{X}(0) = y$ and ending at $\overline{X}(t) = x$.
Then Fujiwara proved, among others, that, for sufficiently small $t > 0$,
the $n\!\rightarrow\! \infty$ limit of the integral kernel $[E(t/n)^n](x,y)$
of $E(t/n)^n$ exists pointwise and is equal to the integral kernel
$e^{-itH_V}(x,y)$ of the Schr\"odinger unitary group $e^{-itH_V}$,
i.e. the fundamental solution for the Schr\"odinger equation,
and further that one has
\begin{equation}%(4.17)
[E(t/n)^n](x,y) - e^{-itH_V}(x,y) = O(n^{-1})t^{2}(2 \pi t)^{-d/2},
\end{equation}
as $n \rightarrow \infty$, uniformly in $x, y$, together with
all the $x, y$--derivatives of the left-hand side, where
$O(n^{-1})$ is independent of $x,\, y$ and $t$.
The proof also yields further that
\begin{equation}%(4.18)
\|E(t/n)^n](x,y) - e^{-itH_V}\|_{L^2\rightarrow L^2} = O(n^{-1}).
\end{equation}
It turns out that this time-sliced approximation to the Schr\"odinger
unitary group $e^{-itH_V}$ converges both pointwise for the integral
kernels and in operator norm, with error estimate $O(n^{-1})$.

Next, in the {\it imaginary-time} case, we will give a little more
detailed account of the related situation than what was briefly mentioned
in \S4.2.1. Assume that $V(x)$ satisfies the condition that there exist
constants $\rho \geq 0$ and $0<\delta \leq 1$ such that
$V(x) \geq C(1+|x|^2)^{\rho/2}$ and
$|\partial_x^{\alpha}V(x)|\leq C_{\alpha}(1+|x|^2)^{(\rho-\delta|\alpha|)/2}$
for every multi-index $\alpha $ with constant $C_{\alpha}$. Here the case
$\delta =0$ is allowed for $\rho = 0$.
Therefore, in particular, it is the case if $V(x)$ is nonnegative
and satisfies the same condition as Fujiwara's. Then the operator
$H_V = -\tfrac12\Delta+V$ becomes selfadjoint with domain
$D[H_V] = D[-\tfrac12\Delta]\cap [V]$. As noted in [I-03],  so
we can obtain analogous results for the Schr\"odinger semigroup $e^{-tH_V}$
with the same error estimate $O(n^{-1})$ in operator norm by the general
abstract theory in [IT-01, ITTZ-01] quoted in \S4.2.1,
and pointwise for the integral kernels as briefly sketched in [I-03, p.86].

However, we have in fact proved much more in [IT-04b, 06] that,
with $F(t) := e^{-\tfrac{t}{2}V}e^{-t\tfrac12(-\Delta)}e^{-\tfrac{t}{2}V}$,
$F(t/n)^n$ converges to $e^{-tH_V}$ with the error estimate $O(n^{-2})$, sharper
than the general optimal $O(n^{-1})$, {\sl both} {\it in operator norm}
{\sl and} {\it pointwise} {\sl for the integral kernels}:
\begin{eqnarray}%(4.19),(4.20)
&&\quad \|F(t/n)^n - e^{-tH_V}\|_{L^2\rightarrow L^2} = O(n^{-2}),\\
&&\quad  [F(t/n)^n](x,y)  - e^{-tH_V}(x,y) = O(n^{-2}) t^{2}(2 \pi t)^{-d/2},
  \,\,\,\, \hbox{\rm uniformly\, on}\,\, {\bf R}^d\times {\bf R}^d,
\end{eqnarray}
locally uniformly in $t>0$.
The error estimate $O(n^{-2})$ here is also seen, in [AzI-08], to be
optimal from below in [AzI-08].
Notice also that this error estimate $O(n^{-2})$ is sharper than
in the real-time case (4.17), (4.18) of the nonrelativistic Schr\"odinger
equation, though the two time-sliced approximations $E(t/n)^n$
and $F(t/n)^n$ are coming from quite different thoughts and ideas.

\bigskip\noindent
{\bf Acknowledgment}. I am most grateful to Professor Daisuke Fujiwara
for illuminating and fruitful discussion on the issue in \S4.2.2 connected
with his works.

%%%%%%%%%%%%%%%%%%%%%%
%{\footnotesize

%}

\end{document}